\documentclass{article}
\usepackage{jheppub}

\usepackage{amsmath,amssymb}

\def \be {\begin{equation}}
\def \ee {\end{equation}}
\def \ba {\begin{array}}
\def \ea {\end{array}}
\def \bea{\begin{eqnarray}}
\def \eea{\end{eqnarray}}

\def \a {\alpha}
\def \b {\beta}
\def \g {\gamma}

\def \G {\Gamma}
\def \d {\delta}

\def \e {\epsilon}

\def \m {\mu}
\def \n {\nu}

\def \l {\lambda}
\def \L {\Lambda}
\def \s {\sigma}
\def \S {\Sigma}
\def \r {\rho}
\def \o {\omega}
\def \O {\Omega}

\def \t {\tau}

\def \p {\partial}
\def \f {\frac}
\def \na {\nabla}

\def \nn {\nonumber}
\def \scl {\ell}
\def \ma {\mathcal}

\def \lt {\left}
\def \rt {\right}

\def \ra {\rightarrow}
\def \lra {\leftrightarrow}
\def \sr {\sqrt}
\def \td {\tilde}
\def \hs {\hspace}

\def \inf {\infty}

\def \cite {\citep}

\def \arcsinh {{\rm arcsinh}}

\def \arctanh {{\rm arctanh}}

\title{\boldmath Aspects of Warped AdS$_3$/CFT$_2$ Correspondence}

\author[a,b,c]{Bin Chen,}
\author[a,b]{Jia-ju Zhang,}
\author[a,b]{Jian-dong Zhang}
\author[a]{ and De-liang Zhong}

\affiliation[a]{Department of Physics, Peking University, Beijing 100871, P.R. China}
\affiliation[b]{State Key Laboratory of Nuclear Physics and Technology, Peking University, Beijing 100871, P.R. China}
\affiliation[c]{Center for High Energy Physics, Peking University, Beijing 100871, P.R. China}

\emailAdd{bchen01@pku.edu.cn}
\emailAdd{jjzhang@pku.edu.cn}
\emailAdd{mksws@pku.edu.cn}
\emailAdd{zdlzdlzdl@gmail.com}

\abstract{In this paper we apply the thermodynamics method to investigate the holographic pictures for the BTZ black hole, the spacelike and the null warped black holes in three-dimensional topologically massive gravity (TMG) and new massive gravity (NMG).  Even though there are higher derivative terms in these theories, the thermodynamics method is still effective. It gives consistent results with the ones obtained by using asymptotical symmetry group (ASG) analysis. In doing the ASG analysis we develop a brute-force realization of the Barnich-Brandt-Compere formalism with Mathematica code, which also allows us to calculate the masses and the angular momenta of the black holes. In particular, we propose the warped AdS$_3$/CFT$_2$ correspondence in the new massive gravity, which states that quantum gravity in the warped spacetime  could  holographically dual to a two-dimensional CFT with $c_R=c_L=\f{24}{Gm\b^2\sr{2(21-4\b^2)}}$.}


\begin{document}

\maketitle

\flushbottom

\section{Introduction}

The three-dimensional (3D) gravity is an important arena for the study of quantum behaviors of gravity. One remarkable feature in  3D Einstein's gravity with a cosmological constant is that even though there is no local dynamical degree of freedom, there are global degrees of freedom. In the seminal work by Brown and Henneaux \cite{Brown:1986nw,Brown:1986ed}, it was shown that under appropriate boundary conditions the asymptotic symmetry group (ASG) of the AdS$_3$ spacetime is generated by two copies of Virasoro algebra with the same central charge $c_R=c_L=\f{3\ell}{2G}$, where $\ell$ is the AdS$_3$ radius and $G$ is 3D Newton constant. In retrospect, this actually present the first example of AdS/CFT correspondence \cite{Maldacena:1997re}. Two important lessons can be drawn from the study in \cite{Brown:1986nw,Brown:1986ed}. One is that there exists boundary degrees of freedom in AdS$_3$. This has been confirmed by the
discovery of three-dimensional BTZ (Banados-Teitelboim-Zanelli) black hole \cite{Banados:1992wn,Banados:1992gq}, and moreover by the successful entropy counting from dual conformal field theory  in \cite{Strominger:1997eq}. The other lesson is that the asymptotic boundary conditions play an essential role in the whole story. One has to find appropriate boundary conditions to obtain a meaningful ASG.

In the past few years, the techniques of ASG analysis has been applied to the study of other 3D gravity theories with higher derivative terms, including 3D topologically massive gravity (TMG) \cite{Deser:1981wh,Deser:1982vy} and 3D new massive gravity (NMG) \cite{Bergshoeff:2009hq,Bergshoeff:2009aq}. The AdS$_3$ spacetime could still be a vacuum, however it turned out that besides the famous Brown-Henneaux boundary conditions there is at least another set of consistent asymptotic boundary conditions leading to dual logarithmic CFT \cite{Grumiller:2008qz}.
Another important feature of these gravities is that there exist other vacua. In the case of 3D TMG, there are spacelike and null warped AdS$_3$ vacua, in which the quantum gravity may have holographical description. This so-called warped
AdS$_3$/CFT$_2$ correspondence was first proposed in \cite{Anninos:2008fx} by studying the holographic description of warped AdS$_3$ black holes in 3D TMG. The ASG analysis was later carried out in \cite{Compere:2008cv,Compere:2009zj,Blagojevic:2009ek}. Further evidence comes from the computations of quasi-normal modes and real-time correlators in \cite{Chen:2009rf,Chen:2009hg,Chen:2009cg,Fareghbal:2010yd,Chen:2010fr,Chen:2010ik,Chen:2010sn}. 
Though there is a relatively good understanding of the warped AdS/CFT correspondence in 3D TMG, it is not clear if the same correspondence could be set up in other 3D gravity theory with more general higher derivatives terms, such as NMG\footnote{After this work was finished, we were informed that the central charge of the warped black holes in NMG has been conjectured in \cite{Kim:2009jm} using the entropy function method \cite{Sen:2005wa}}. This is one of motivations of this work.

Another important application of ASG analysis is in the Kerr/CFT correspondence \cite{Guica:2008mu} and its extensions. In these cases, the asymptotically flat black hole with angular momentum or U(1) charges could be holographically described by a 2D CFT. To read the central charges of dual CFT, one has to do ASG analysis on the near-horizon geometry of extremal black hole. This near horizon geometry is  similar to a warped spacetime, but the asymptotic boundary conditions are very different. Strictly speaking,  the ASG analysis can only be applied to the extremal black holes, though the correspondences have been extended to non-extremal cases with the help of hidden conformal symmetry in the low frequency scattering off the black hole \cite{Castro:2010fd}. It is still an open issue how to read the central charges of dual CFT for generic non-extremal black holes.

Very recently, the black hole/CFT (BH/CFT) correspondence was investigated from the point of view of thermodynamics of both the outer and inner horizons \cite{Chen:2012mh}. The key point is that the inner horizon thermodynamics may play an essential role in setting up the BH/CFT correspondence\footnote{The first law of the inner horizon thermodynamics takes the form as $dM=-T_-dS_- + \cdots$. In fact it is not precise to call it as the first law of thermodynamics in the usual sense. A better way to understand it is that such a relation reflects how the inner horizon response to a perturbation carrying mass, angular momentum or charges. Here we still use ``thermodynamics law" for this relation, as widely  used in the community.}. From the thermodynamics laws of both horizons, it is straightforward to read the thermodynamics of left- and right-mover and the corresponding dual temperatures. This method has been
tested and applied to many  cases \cite{Chen:2012mh,Chen:2012yd,Chen:2012ps,Chen:2012pt,Chen:2013rb}. Especially in \cite{Chen:2013rb}, the relation between the thermodynamics method and other conventional methods has been clarified. It turns out that they are consistent with each other for general black holes in Einstein(-Maxwell) theory and in some cases the thermodynamics method is even more powerful. Up till now this method has only been  used for the black holes in the Einstein(-Maxwell) gravity. This class of gravity  theory is diffeomorphism invariant, and the dual two-dimensional CFT is required to have $c_R=c_L$. In the thermodynamics method, $c_R=c_L$ is equivalent to  the condition $T_+S_+=T_-S_-$ with $T_\pm, S_\pm$ being the outer and inner horizon temperatures and entropies respectively, or equivalently the condition that the entropy product $S_+S_-$ being mass-independent. Therefore for the black holes in the Einstein gravity and other diffeomorphism invariant gravity theories, the condition $T_+S_+=T_-S_-$ may be taken as the criterion whether there is a 2D CFT dual. But for a black hole in a gravity theory which has diffeomorphism anomaly, like 3D TMG, $c_R \neq c_L$ is expected, so $T_+S_+=T_-S_-$ cannot be the criterion for the existence of the CFT dual. It is interesting to see if the thermodynamics method can be applied to this case.  Moreover it is also important to know if the thermodynamics method could be applied  for the black holes in a gravity theory with higher curvature terms. In the present work  we would like to investigate how to apply thermodynamics method in 3D TMG and other gravity theory with high curvature terms, taking 3D NMG as prototype.
More precisely, we shall focus on the holographic pictures for the warped black holes in these theories.

The remaining parts are arranged as follows. In Section~\ref{s2} we give an introduction to 3D gravities, including the Einstein gravity with a negative cosmological constant, TMG and NMG. We also show how to define various thermodynamical quantities in these theories. In Section~\ref{s3} we give a brief review of thermodynamics methods in setting up holographic pictures for the black holes. In Section~\ref{s4} we consider the BTZ black hole in GR, TMG and NMG, as tests of the validity of the thermodynamics method and the Barnich-Brandt-Compere (BBC) formalism. In Section~\ref{s5} and \ref{s6}, we discuss the CFT duals of the spacelike and null warped black holes in both TMG and NMG, from the points of view of thermodynamics and ASG analysis. We end with conclusion and discussion in Section~\ref{s7}. In the Appendix \ref{sa}, we review the BBC formalism and present a brute-force realization for it using Mathematica code.

\section{Thermodynamics of black holes in GR, TMG and NMG}\label{s2}

In this section we briefly introduce 3D general relativity with a negative cosmological constant (GR), topologically massive gravity (TMG) and new massive gravity (NMG). Especially we give the formulas of calculating the thermodynamics quantities for general black holes in GR, TMG and NMG.

\subsection{GR}

For GR, the action  and the equation of motion are respectively
\bea \label{gr}
&& I_{GR}=\f{1}{16\pi G}\int d^3x \sr{-g}(R-2\L),  \nn\\
&& R_{\m\n}-\f{1}{2}R g_{\m\n}+\L g_{\m\n}=0.
\eea
with $G$ being the Newton constant and $\L$ being the cosmological constant.

Given a three-dimensional black hole with the outer and inner horizons $r_\pm$, we could write the metric of the black hole in a ADM form
\be \label{bh}
ds^2=-N^2 dt^2+g_{rr}dr^2+g_{\phi\phi}\lt( d\phi+N^\phi dt \rt)^2,
\ee
with $\phi \sim \phi +2\pi$. We require that the black hole has isometries $\p_t$ and $\p_\phi$, and then we could always let $N^\phi|_{r \to \inf}=0$ by possible redefinition of $\phi \to \phi -N^\phi|_{r \to \inf} t$. The Hawking temperatures and the angular velocities of the outer and inner horizons are
\bea \label{to}
&& T_\pm=\lt| \f{\p_r N^2}{4\pi\sr{N^2 g_{rr}}} \rt|_{r=r_\pm}, \hs{3ex}\O_\pm=-N^\phi|_{r=r_\pm}.
\eea
Note that these thermodynamical quantities depend only on the geometry of the black hole and are independent of the theory the black hole belong to. In other words the formulas in (\ref{to}) apply not only to the black holes in GR, but also to the black holes in TMG and NMG. For a black hole in GR the entropy is just Bekenstein-Hawking entropy, one quarter of the horizon area in Planck unit. Then the outer and inner horizon entropies of the three-dimensional black hole (\ref{bh}) are
\be \label{sgr}
S_\pm=\lt.  \f{\pi \sr{g_{\phi\phi}}}{2G}  \rt|_{r=r_\pm}.
\ee

Generically, the black hole (\ref{bh}) is characterized by two parameters,  the mass $M$ and the angular momentum $J$, or equivalently the parameters $r_\pm$ characterizing the outer and inner horizons, or some other independent two parameters. There are more than one ways of calculating the mass and the angular momentum of the black hole, and in the present work we will use the brute-force realization of the BBC formalism presented in the appendix. In the BBC formalism, the mass and the angular momentum are both called charges. When the parameters of the black hole are $r_\pm$, the charge difference between the black hole $g_{\mu\nu}$ (\ref{bh}) and its variation $g_{\m\n}+\d g_{\m\n}$ with
\be
\d g_{\m\n}=\f{\p g_{\m\n}}{\p r_+} d r_+  +  \f{\p g_{\m\n}}{\p r_-} d r_-,
\ee
is defined as (\ref{charge}) $Q_\xi[g+\d g;g]$. With $\xi=\p_t$, we have
\be
dM=\f{\p M}{\p r_+} d r_+  +  \f{\p M}{\p r_-} d r_-=Q_{\p_t}.
\ee
The mass $M$ could be obtained from $Q_{\p_t}$ by integrating from $(0,0)$ to $(r_+,r_-)$. Similarly one could get the angular momentum of the black hole using $dJ=-Q_{\p_\phi}$. When the parameters of the black hole are not $r_\pm$ the process is similar, and one could see more details in the Mathematica code attached to this paper. Having got the thermodynamic quantities at both the outer and inner horizons, we usually have the first laws at the outer and inner horizons
\bea
&&d M=T_+ d S_+ + \O_+d J  \nn\\
&&\phantom{d M}=-T_- d S_- + \O_-d J.
\eea
There are the same first laws for black holes in TMG and NMG, and we will not bother to repeat them in the following two subsections.

\subsection{TMG}

The TMG was proposed in \cite{Deser:1981wh,Deser:1982vy} and its action and equation of motion are respectively
\bea \label{tmg}
&& I_{TMG}=\f{1}{16\pi G}\int d^3x \sr{-g} \lt[ R-2\L  +\f{1}{2\m}\e^{\l\m\n}\G^\r_{\l\s}
                             \lt( \p_\m\G^\s_{\r\n} +\f{2}{3}\G^\s_{\m\t}\G^\t_{\n\r} \rt) \rt],   \nn\\
&& R_{\m\n}-\f{1}{2}R g_{\m\n}+\L g_{\m\n}+\f{1}{\m}C_{\m\n}=0,
\eea
with $\m$ being the coupling constant, $\e^{\l\m\n}$ being the Levi-Civita tensor and $C_{\m\n}$ being the Cotton tensor
\be
C_{\m\n}=\e_{\m}^{\phantom{\m}\r\s}\na_\r \lt( R_{\s\n}-\f{1}{4}g_{\s\n}R \rt).
\ee
Note that there are at most third derivatives in the Cotton tensor. We choose the convention $\e^{tr\phi}=\f{-1}{\sr{-g}}$ when we use the coordinates $(t,r,\phi)$.

If the black hole (\ref{bh}) is a solution of TMG, its Hawking temperatures $T_\pm$ and angular velocities $\O_\pm$ are the same as (\ref{to}). The mass $M$ and the angular momentum $J$ of the black hole could also be calculated using the BBC formalism similar to the procedure for the black hole in GR, but as the action of the theory has changed, so the superpotential and the charges have to be modified accordingly.
One should be careful to calculate the entropies of the black holes in TMG, because the higher order derivative terms in the action (\ref{tmg}) have contribution to the entropy. To get the entropies of black holes in a higher derivative gravity, one can use Wald formula \cite{Wald:1993nt,Jacobson:1993vj,Iyer:1994ys} or equivalently the conical singularity method \cite{Solodukhin:1994yz,Fursaev:1995ef}. It can be shown that using the results in \cite{Solodukhin:2005ah,Tachikawa:2006sz} for a black hole in TMG (\ref{tmg}) with the metric (\ref{bh}) the entropies at the outer and inner horizons are \cite{Chen:2011dc}
\be \label{stmg}
S_\pm= \lt. \f{\pi \sr{g_{\phi\phi}}}{2G}+\f{\pi g_{\phi\phi}\p_r N^\phi}{4G\m\sr{N^2 g_{rr}}} \rt|_{r=r_\pm}.
\ee

\subsection{NMG}

The NMG was proposed in \cite{Bergshoeff:2009hq,Bergshoeff:2009aq}, and its action is
\be \label{nmg}
I_{NMG}=\f{1}{16\pi G}\int d^3x \sr{-g} \lt( R-2\l-\f{1}{m^2}K \rt),
\ee
with $m$ being the coupling constant and
\be
K=R_{\m\n}R^{\m\n}-\f{3}{8}R^2.
\ee
We just suppose $m>0$ for simplicity and do not consider the possible analytical continuation.  Due to the existence of the $K$ term, $\l$ is not necessarily the cosmological constant. We define the three-dimensional cosmological constant $\L$ as $R=6\L$. The equation of motion for NMG is
\be
R_{\m\n}-\f{1}{2}R g_{\m\n}+\l g_{\m\n}-\f{1}{2m^2}K_{\m\n}=0,
\ee
with
\be
K_{\m\n}=-\f{1}{2}\na^2R g_{\m\n}-\f{1}{2}\na_\m \na_\n R+2\na^2R_{\m\n}-8R_\m^{\phantom{\m}\a}R_{\a\n}+\f{9}{2}R R_{\m\n}
         +g_{\m\n} \lt( 3R_{\a\b}R^{\a\b}-\f{13}{8}R^2  \rt).
\ee
The tensor $K_{\m\n}$ involves the fourth derivative of the metric. In deriving  $K_{\m\n}$, we have used the fact that the Weyl tensor is always vanishing in three dimensions, or equivalently the identity
\be
R_{\r\s\m\n}=g_{\r\m}R_{\s\n}+g_{\s\n}R_{\r\m}-g_{\r\n}R_{\s\m}-g_{\s\m}R_{\r\n}-\f{R}{2}(g_{\r\m}g_{\s\n}-g_{\r\n}g_{\s\m}).
\ee

The mass $M$ and the angular momentum $J$ of a black hole (\ref{bh}) in NMG (\ref{nmg}) can be calculated using the BBC formalism. The entropies at the outer and inner horizons could be calculated using conical singularity  method \cite{Solodukhin:1994yz,Fursaev:1995ef} as
\be \label{snmg}
S_\pm= \lt. \f{\pi \sr{g_{\phi\phi}}}{2G} \lt[ 1-\f{1}{m^2}( R_{ii}-\f{3}{4}R  ) \rt] \rt|_{r=r_\pm},
\ee
with
\bea
&& R_{ii}=-R_{\m\n}n_1^\m n_1^\n+R_{\m\n}n_2^\m n_2^\n,  \nn\\
&& n_1^\m=\f{1}{N}(1,0,-N^\phi),  \nn\\
&& n_2^\m=(0,\f{1}{\sr{g_{rr}}},0).
\eea

\section{Thermodynamics method of black hole/CFT correspondence}\label{s3}

The thermodynamics method was proposed and developed in \cite{Chen:2012mh,Chen:2012yd,Chen:2012ps,Chen:2012pt,Chen:2013rb}. In \cite{Chen:2013rb} there is a systematic summary of the method. For a nonextremal black hole, one could get much universal information of dual CFT, including the right- and left-moving central charges and temperatures, from the thermodynamics laws of the outer and inner horizons.  Here we only review the main conclusions of the method, and one could find the details in the papers cited above.

Suppose that the black hole we are interested in has only two physical horizons. If the
black hole is stationary, then one can prove that there are  first laws of thermodynamics for both the outer and inner horizons in the Einstein(-Maxwell) gravity. In the present work, we focus on the 3D black holes without charge or scalar hair. In this case, the black hole is characterized by the mass and angular momentum. The first laws turn out to be
\bea \label{law1}
&&d M=T_+ d S_+ + \O_+d J  \nn\\
&&\phantom{d M}=-T_- d S_- + \O_-d J.
\eea
We could recombine these quantities into the ones of separated left- and right-moving sectors
\bea \label{def}
&& T_{R,L}=\f{T_-T_+}{T_- \pm T_+},
~~~ S_{R,L}=\f{1}{2}(S_+ \mp S_-),  \nn\\
&& \O_{R}=\f{T_- \O_+ + T_+ \O_-}{2(T_-+T_+)},
~~~ \O_{L}=\f{T_- \O_+ - T_+ \O_-}{2(T_--T_+)},
\eea
such that
\bea \label{law2}
&&\f{1}{2}d M=T_R d S_R+\O_Rd J  \nn\\
&&\phantom{\f{1}{2}d M}=T_L d S_L+\O_L d J.
\eea

Let us define
\be \label{rj}
R_J=\f{1}{\O_R-\O_L}=\f{T_-^2-T_+^2}{T_- T_+(\O_- - \O_+)},
\ee
which is just the scale of the space that the two dimensional CFT resides. From (\ref{law2}) we get
\be \label{law3}
dJ=T_L^J d S_L-T_R^J d S_R,
\ee
with the dimensionless temperatures of CFT dual to the black hole being
\be \label{tt}
T^J_{R,L}=R_J T_{R,L}=\f{T_- \mp T_+}{\O_--\O_+}.
\ee
From the Cardy formula we could derive the central charges
\be \label{cc1}
c_{R,L}^J=\f{3}{\pi^2}\f{S_{R,L}}{T_{R,L}^J}
=\f{3}{2\pi^2}\f{(\O_--\O_+)(S_+ \mp S_-)}{T_- \mp T_+}.
\ee

In the above discussion, we only used the first laws of thermodynamics no matter what kind of theory the black hole belong to. For a diffeomorphism invariant theory, like GR and NMG, there should be $c_R=c_L$ for the dual CFT.
From (\ref{cc1}) it can be shown easily that $c_R=c_L$ is equivalent to $T_+ S_+=T_- S_-$. According to the first laws (\ref{law1}) this is also equivalent to the condition that the entropy product $S_+S_-$ is mass-independent. However, in TMG there is diffeomorphism anomaly, and therefore it is expected that $c_R \neq c_L$ for the dual CFT. Consequently, the entropy product $S_+S_-$ is mass-dependent in TMG, as has been observed in \cite{Detournay:2012ug}. Nevertheless the above treatment still makes sense without trouble. Actually one may check if it is consist with the diffeomorphism anomaly.

In fact more information could be read from the thermodynamics \cite{Chen:2012ps,Chen:2013rb}. In the gravity side we throw the perturbation $dM=\o, dJ=k$ into the black hole, and rewrite the first laws (\ref{law2})  as
\bea
&& T_R^J d S_R= R_J \lt( \f{1}{2}\o-\O_R k \rt),  \nn\\
&& T_L^J d S_L= R_J \lt( \f{1}{2}\o-\O_L k \rt).
\eea
In the CFT side, we suppose that there are
\bea
&& T_R^J d S_R=\o_R^J-q_R^J \m_R^J,  \nn\\
&& T_L^J d S_L=\o_L^J-q_L^J \m_L^J,
\eea
with $\o_{R,L}^J$, $q_{R,L}^J$ and $\m_{R,L}^J$ being the frequencies, the charges, and the  chemical potentials of the corresponding operator perturbing the thermal equilibrium. Then we find the identifications
\bea
&& \o_{R,L}^J=\f{R_J}{2}\o=\f{T_-^2-T_+^2}{2T_- T_+(\O_- - \O_+)}\o,  \nn\\
&& q_{R,L}^J=k,  \nn\\
&& \m_{R}^J=R_J \O_{R}=\f{(T_- - T_+)(T_-\O_+ + T_+\O_-)}{2T_-T_+(\O_- - \O_+)},  \nn\\
&& \m_{L}^J=R_J \O_{L}=\f{(T_- + T_+)(T_-\O_+ - T_+\O_-)}{2T_-T_+(\O_- - \O_+)}.
\eea
Notice that these quantities of the dual CFT depend only on the the geometry of the black hole, and have nothing to do with what theory the black hole belong to. They are always the same as the ones found in the low frequency scattering amplitude as shown in \cite{Chen:2013rb}. This is sensible because the scattering amplitude also depends only on the the geometry of the black hole.

\section{BTZ black hole} \label{s4}

In this section we consider the three-dimensional BTZ  black hole in GR, TMG and NMG. We have to say that all the results got in this section are not new. The central charges of AdS$_3$ spacetime and the statistical derivation of the entropy of BTZ black hole in GR have been given long ago in \cite{Brown:1986nw,Brown:1986ed,Strominger:1997eq,Balasubramanian:1999re}. The central charges of AdS$_3$, the mass, the angular momentum and the entropy of BTZ black hole
in TMG, NMG and other high curvature gravity have been discussed  in \cite{Saida:1999ec,Garcia:2003nm,Moussa:2003fc,Deser:2003vh,Deser:2005jf,Olmez:2005by,Kraus:2005zm,Solodukhin:2005ah,Sahoo:2006vz,
Park:2006gt,Park:2006zw,Tachikawa:2006sz,Bouchareb:2007yx,Hotta:2008yq,Clement:2009gq,Liu:2009bk}. However, we take this case as a warmup, firstly to show how the thermodynamics method in setting up BH/CFT correspondence applies to BTZ black hole in TMG and NMG, and secondly to test our code of calculating the charges and central charges in GR, TMG and NMG.

\subsection{Black hole solution}

The metric of BTZ black hole could be written as
\be \label{btz}
ds^2=-\f{(r^2-r_+^2)(r^2-r_-^2)}{\ell^2r^2}dt^2+\f{\ell^2r^2}{(r^2-r_+^2)(r^2-r_-^2)}dr^2+r^2 \lt( d\phi-\f{r_+r_-}{\ell r^2}dt \rt)^2,
\ee
with $\ell>0$ being the parameter of the theory and $r_- \leq r_+$ being the horizons of the black hole. The Hawking temperatures and the angular velocities at the outer and inner horizons are respectively
\bea \label{btzto}
&& T_\pm=\f{r_+^2-r_-^2}{2\pi\ell^2 r_\pm},  \nn\\
&& \O_+=\f{r_-}{\ell r_+}, ~~~ \O_-=\f{r_+}{\ell r_-}.
\eea

The BTZ black hole (\ref{btz}) is the solution of GR (\ref{gr}) with $\L=-1/\ell^2$. From the BBC formalism one could get the mass and the angular momentum of the black hole as
\be
M=\f{r_+^2+r_-^2}{8G\scl^2}, ~~~ J=\f{r_+r_-}{4G\scl}.
\ee
From (\ref{sgr}) one get the entropies at the outer and inner horizons
\be
S_\pm=\f{\pi r_\pm}{2G}.
\ee

The BTZ black hole (\ref{btz}) is also the solution of TMG (\ref{tmg})  with $\L=-1/\ell^2$. The mass and the angular momentum could be calculated by the BBC formalism as
\be
M=\f{1}{8G\ell^2} \lt( r_+^2+r_-^2 +\f{1}{\m\ell}2r_+r_- \rt), ~~~
J=\f{1}{4G\ell} \lt( r_+r_- +\f{1}{\m\ell}\f{r_+^2+r_-^2}{2} \rt).
\ee
 From the relation (\ref{stmg})  the outer and inner horizons entropies are respectively
\be
S_+=\f{\pi}{2G} \lt( r_+ + \f{1}{\m\ell}r_- \rt),  ~~~
S_-=\f{\pi}{2G} \lt( r_- + \f{1}{\m\ell}r_+ \rt).
\ee
As being checked in \cite{Detournay:2012ug}, for the BTZ black hole in TMG the entropy product $S_+ S_-$ is mass-dependent. For physical reasons we require that
\be \label{e22}
M \geq 0, ~~~ 0 \leq S_- \leq S_+
\ee
are satisfied for all values of $0 \leq r_- \leq r_+$ and this requires that $\m\ell \geq 1$.

The BTZ black hole is also the solution of NMG (\ref{nmg}) with $\l=\L-\f{\L^2}{4m^2}, \L=-\f{1}{\ell^2}$. The mass and the angular momentum of BTZ black hole are respectively
\be
M=\f{r_+^2+r_-^2}{8G\scl^2}\lt( 1-\f{1}{2m^2\scl^2} \rt), ~~~
J=\f{r_+r_-}{4G\scl} \lt( 1-\f{1}{2m^2\scl^2} \rt).
\ee
From (\ref{snmg}), one could also get
\be
S_\pm=\f{\pi r_\pm}{2G} \lt( 1-\f{1}{2m^2\ell^2} \rt).
\ee
Since NMG is a diffeomorphism invariant theory, we must have $c_R=c_L$, and so $T_+S_+=T_-S_-$ which can be verified easily. The requirements $M \geq 0,0 \leq S_- \leq S_+$ imply that $m^2\ell^2 \geq \f{1}{2}$.

\subsection{CFT from thermodynamics}

For the BTZ black hole in GR, TMG or NMG, there are always the first laws (\ref{law1}) of the outer and inner horizons. From the quantities (\ref{btzto}), we could get the scale of the space the CFT resides and the temperatures of the CFT as
\be
R_J=\ell, ~~~ T_{R,L}^J=\f{r_+ \mp r_-}{2\pi\ell}.
\ee

In GR, from (\ref{def}) and (\ref{cc1}) the right- and left-moving entropies of the dual CFT are
\be
S_{R,L}=\f{\pi (r_+ \mp r_-)}{4G},
\ee
and then the central charges are
\be
c_{R,L}^J=\f{3\ell}{2G}.
\ee

In TMG, the right- and left-moving entropies of the dual CFT are
\be
S_{R}=\f{\pi (r_+ - r_-)}{4G} \lt( 1-\f{1}{\m\ell} \rt), ~~~
S_{L}=\f{\pi (r_+ + r_-)}{4G} \lt( 1+\f{1}{\m\ell} \rt),
\ee
and then the central charges are
\be
c_{R,L}^J=\f{3\ell}{2G} \lt( 1 \mp \f{1}{\m\ell} \rt).
\ee
Note that the requirement guarantees that $0\leq S_R \leq S_L$ and $c_{R,L}^J \geq 0$.
When $\m\ell=1$, the theory becomes chiral with vanishing right central charge.

In NMG, the entropies and central charges of the dual CFT are respectively
\bea
&& S_{R,L}=\f{\pi (r_+ \mp r_-)}{4G} \lt( 1-\f{1}{2m^2\ell^2} \rt), \nn\\
&& c_{R,L}^J=\f{3\ell}{2G}\lt( 1-\f{1}{2m^2\ell^2} \rt).
\eea

\subsection{CFT from ASG}

The central charges of AdS$_3$ in GR, TMG and NMG could be obtained by using the brute-force realization of the BBC formalism presented in Appendix~\ref{sa}. Setting $r_+=r_-=0$ in the metric of the BTZ black hole, one could get the metric of the AdS$_3$ spacetime
\be \label{ads3}
ds^2=-\f{r^2}{\ell^2} dt^2+\f{\ell^2}{r^2}dr^2+r^2d\phi^2.
\ee
Using the Brown-Henneaux boundary conditions \cite{Brown:1986nw,Strominger:1997eq} in $(t,r,\phi)$ coordinates for AdS$_3$
\be
\d g_{\m\n}=\ma O \lt(
\ba{ccc}
1&\f{1}{r^3}&1\\
&\f{1}{r^4}&\f{1}{r^3}\\
&&1
\ea \rt),
\ee
one could have the asymptotic Killing vector $\xi=\xi^\m\p_\m$ with leading terms
\bea
&& \xi^t=\ell (T^+ + T^-) + \f{\ell^3}{2r^2}(\p_+^2 T^+ + \p_-^2 T^-), \nn\\
&& \xi^r=-r(\p_+ T^+ + \p_- T^-),  \nn\\
&& \xi^\phi=T^+ - T^- - \f{\ell^2}{2r^2}(\p_+^2 T^+ - \p_-^2 T^-),
\eea
where $x^\pm=\f{t}{\ell} \pm \phi, 2\p_\pm=\ell\p_t \pm\p_\phi$, and $T^\pm=T^\pm(x^\pm)$. Expanding $T^\pm=\f{1}{2}e^{imx^\pm}$, one could get the ASG which forms the Witt algebras through Lie brackets,
\bea
&& i[\xi_m^\pm,\xi_n^\pm]=(m-n)\xi_{m+n}^\pm,  \nn\\
&& [\xi_m^+,\xi_n^-]=0.
\eea

 Using the Mathematica code we find that in GR, the central charges are
\be
c_{R,L}=\f{3\scl}{2G},
\ee
in TMG they are
\be
c_{R,L}=\f{3\scl}{2G}\lt(1 \mp \f{1}{\m\ell}\rt),
\ee
and in NMG they are
\be
c_{R,L}=\f{3\scl}{2G}\lt(1-\f{1}{2m^2\scl^2}\rt).
\ee
These central charges are the same as the ones got from the thermodynamics method before.

\section{Spacelike warped black hole} \label{s5}

In this section we consider the spacelike warped black hole constructed in \cite{Moussa:2003fc,Bouchareb:2007yx,Moussa:2008sj,Clement:2009gq}. The black hole is not the solution of GR but could be the solution of TMG and NMG. 

\subsection{Black hole solution}

Let us start from the spacelike warped black hole in NMG. It  has the metric
\be \label{swbh}
ds^2=-\b^2\f{r^2-r_0^2}{R(r)^2}dt^2+\f{dr^2}{\g^2(r^2-r_0^2)}+R(r^2)\lt[ d\phi-\f{r+(1-\b^2)\O}{R(r)^2}dt \rt]^2
\ee
with
\be
R(r)^2=r^2+2\O r+(1-\b^2)\O^2+\f{\b^2r_0^2}{1-\b^2}.
\ee
For the metric (\ref{swbh}) to be the solution of NMG (\ref{nmg}), we should have
\bea
&& \b^2=\f{63+2m^2\ell^2}{4(3+2m^2\ell^2)}, \nn\\
&& \g^2=\f{63+2m^2\ell^2}{20\ell^2},  \nn\\
&& \l=\f{189-468m^2\ell^2+4m^4\ell^4}{400m^2\ell^4}.
\eea
Note that besides the Newton constant $G$ there are two parameters $(\l,m)$ for the NMG theory (\ref{nmg}), and one could represent them by $(m,\ell)$ or some other pairs. In this section we use the pair $(\b,m)$ as independent parameters of the theory for simplicity. The other parameters could be represented by them
\bea \label{e33}
&& \g^2=\f{8m^2\b^2}{21-4\b^2},  \nn\\
&& \l=\f{m^2(21-72\b^2+16\b^4)}{(21-4\b^2)^2}.
\eea
Here we suppose $\b,\g>0$ and do not try to make any analytical continuation. The other two parameters $r_0>0,\O$ in the solution characterize the black hole. For the black hole to be causally regular and geodecically complete there should be conditions \cite{Bouchareb:2007yx,Moussa:2008sj}
\be
0<\b^2<1, ~~~  \O \geq -\f{r_0}{\sr {1-\b^2}} ~ {\rm and} ~ \O \neq \f{r_0}{1-\b^2}.
\ee
For later convenience when $-\f{r_0}{\sr {1-\b^2}} \leq \O <\f{r_0}{1-\b^2}$ we call the black hole as slow-rotating and when $\O > \f{r_0}{1-\b^2}$, we call it fast-rotating  black hole. As will be seen below, when $r_0$ is fixed a fast-rotating black hole can have the angular momentum as large as possible, and for a slow-rotating one the absolute value of the angular momentum has a upper bound. Note that the two regions cannot be connected by continuously variations of the parameters.

 The spacetime (\ref{swbh}) has no constant Ricci tensor but  has a constant Ricci scalar
\be
R=-\f{(4\b^2-1)\g^2}{2\b^2}.
\ee
For the black hole being a warped AdS black hole, we need $R=-\f{6}{\ell^2}<0$, which implies that
\be \label{e31}
\f{1}{4} < \b^2 <1, ~~~ \g^2=\f{12\b^2}{\scl^2(4\b^2-1)}.
\ee

The spacelike warped black hole has outer and inner horizons at $r= \pm r_0$. For a fast-rotating black hole, we could calculate the Hawking temperatures, the angular velocity and the entropies at the outer and inner horizons
\bea
T_\pm&=&\f{\b\g\sr{1-\b^2}r_0}{2\pi[(1-\b^2)\O \pm r_0]}, ~~~ \label{e25}\\ \O_\pm&=&\f{1-\b^2}{(1-\b^2)\O \pm r_0},\nn\\
S_\pm&=&\f{8\pi\lt[(1-\b^2)\O \pm r_0\rt]}{G\sr{1-\b^2}(21-4\b^2)}.\label{e29}
\eea
For a slow-rotating black hole, the angular velocity does not change but the Hawking temperatures and the entropies become
\bea
T_\pm&=&\f{\b\g\sr{1-\b^2}r_0}{2\pi[r_0 \pm (1-\b^2)\O]}, \label{e26} \\
S_\pm&=&\f{8\pi\lt[r_0 \pm (1-\b^2)\O\rt]}{G\sr{1-\b^2}(21-4\b^2)}.\label{e30}
\eea
Using the BBC formalism we get the mass and the angular momentum
\bea
&& M=\f{8\sr{2}m\b^2(1-\b^2)\O}{G(21-4\b^2)^{3/2}}, \nn\\
&& J=\f{4\sr{2}m\b^2\lt[ (1-\b^2)^2\O^2 - r_0^2 \rt]}{G(1-\b^2)(21-4\b^2)^{3/2}}.
\eea
One could check that the first laws (\ref{law1}) hold for both the fast and slow-rotating black holes.
The requirements $M>0$ and $0\leq S_- \leq S_+$ are satisfied for the fast-rotating black hole, but give an upper bound on the angular velocity for the slow-rotating black hole
\be
0 \leq \O <\f{r_0}{1-\b^2}.
\ee

The spacetime (\ref{swbh}) could also be the solution of TMG (\ref{tmg}) by the following identification of the parameters
\be
\b^2=\f{\n^2+3}{4\n^2}, ~~~ \g^2=\f{\n^2+3}{\ell^2},
\ee
with $\L=-\f{1}{\scl^2}$, $\m=-\f{3\n}{\scl}$ and $\n>1$. As the higher derivative terms in TMG are different from the ones in NMG, the entropies, the mass and the angular momentum get modified. For a fast-rotating black hole the entropies of the outer and inner horizons could be calculated using (\ref{stmg})
\be \label{e27}
S_\pm=\f{\pi\lt[ 3(\n^2-1)\O \pm (5\n^2+3)r_0 \rt]}{6G\n\sr{3(\n^2-1)}},
\ee
and for a slow-rotating black hole they become
\be \label{e28}
S_\pm=\f{\pi\lt[ (5\n^2+3)r_0 \pm 3(\n^2-1)\O  \rt]}{6G\n\sr{3(\n^2-1)}}.
\ee
Using the BBC formalism we could get the mass and the angular momentum as
\bea
&& M=\f{(\n^2-1)(\n^2+3)\O}{16G\ell\n^3}, \nn\\
&& J=\f{(\n^2+3)\lt[ 9(\n^2-1)^2\O^2-4\n^2(5\n^2+3)r_0^2 \rt]}{288G\ell\n^3(\n^2-1)}.
\eea
We require that $M\geq0$ and $0\leq S_- \leq S_+$, and this is satisfied for the fast-rotating black hole, but for the slow-rotating black hole there is additional constraint
on the angular velocity
\be
0 \leq \O <\f{r_0}{1-\b^2}=\f{4\n^2 r_0^2}{3(\n^2-1)}.
\ee
The first laws (\ref{law1}) are satisfied for both the fast and slow-rotating black holes.

\subsection{CFT from thermodynamics}

For the warped black hole, we first consider the fast-rotating black hole with $\O>\f{r_0}{1-\b^2}$.
From the quantities (\ref{e25}), we could get the scale of the space the CFT resides and the temperatures of the CFT as
\be \label{e2}
R_J=2\O,~~~
T_R^J=\f{2m\b^2 r_0}{\pi \sr{2(1-\b^2)(21-4\b^2)}}, ~~~
T_L^J=\f{m\b^2\O}{\pi} \sr{\f{2(1-\b^2)}{21-4\b^2}}.
\ee
For a perturbation $dM=\o,dJ=k$ of the black hole, the dual operator in the CFT has frequencies, charges and chemical potentials
\be
\o_{R,L}^J=\O \o, ~~~ q_{R,L}^J=k, ~~~ \m_{R}^J=1, ~~~ \m_{L}^J=0.
\ee
Considering the coordinate transformations presented in \cite{Anninos:2008fx}, we find that those quantities are consistent with the ones in \cite{Fareghbal:2010yd}. In TMG, the right- and left-moving entropies, temperatures and  central charges of the dual CFT are respectively
\bea
&& S_R=\f{\pi(5\n^2+3)r_0}{6G\n\sr{3(\n^2-1)}},~~~
S_L=\f{\pi\O}{2G\n}\sr{\f{\n^2-1}{3}},   \nn\\
&& T_R^J=\f{(\n^2+3)r_0}{2\pi\ell\sr{3(\n^2-1)}}, ~~~
T_L^J=\f{(\n^2+3)\sr{3(\n^2-1)}\O}{8\pi\n^2\ell},  \nn\\
&& c_R^J=\f{(5\n^2+3)\ell}{G\n(\n^2+3)}, ~~~
c_L^J=\f{4\ell\n}{G(\n^2+3)}.
\eea
The central charges are the same as proposed in \cite{Anninos:2008fx}.

In NMG, these quantities are respectively
\bea
&& S_R=\f{8\pi r_0}{G(21-4\b^2)\sr{1-\b^2}},
~~~ S_L=\f{8\pi \sr{1-\b^2}\O}{G(21-4\b^2)},    \nn\\
&& T_R^J=\f{2m\b^2 r_0}{\pi \sr{2(1-\b^2)(21-4\b^2)}},
~~~ T_L^J=\f{m\b^2\O}{\pi} \sr{\f{2(1-\b^2)}{21-4\b^2}},  \nn\\
&& c_{R,L}^J=\f{24}{Gm\b^2\sr{2(21-4\b^2)}}. \label{warpNMG}
\eea
The central charges are just the ones in \cite{Kim:2009jm}.

For the slow-rotating black hole with $0\leq \O< \f{r_0}{1-\b^2}$, because $\O_--\O_+<0$ one has to redefine (\ref{rj})
\be
R_J=\f{T_-^2-T_+^2}{T_- T_+(\O_+ - \O_-)},
\ee
and consequently one has
\be
dJ=T_R^J dS_R-T_L^J dS_L.
\ee
From (\ref{e27}), (\ref{e28}), (\ref{e29}) and (\ref{e30}), comparing with the fast-rotating black hole, the outer horizon entropy of the slow-rotating black hole does not change but the inner horizon entropy gets a minus sign. Thus $S_{R,L}$ for the fast-rotating black hole become $S_{L,R}$ for the slow-rotating one, and the same things happen to other quantities. Therefore the right- and left-moving sectors of the CFT dual to the fast-rotating black hole become the left- and right-moving sectors of the CFT dual to the slow-rotating one.

\subsection{CFT from quotient}

In this section we follow \cite{Anninos:2008fx} and show that the CFT temperatures (\ref{e2}) of the spacelike warped AdS$_3$ black hole (\ref{swbh}) could be obtained by identifying the black hole as the discrete quotient of the spacelike warped AdS$_3$ spacetime. The metric of the spacelike warped AdS$_3$ spacetime is
\be \label{wads3}
ds^2=\f{1}{\g^2} \lt[ -\cosh^2\s d\t^2 +d\s^2 +\f{1}{\b^2}(du+\sinh\s d\t)^2 \rt],
\ee
with the constraints (\ref{e31}). The geometry has SL(2,$R$)$\times$U(1) isometries.   The explicit forms of corresponding  Killing vectors $\{\td J_0,\td J_1,\td J_2\}$ and $J_2$ could be found in \cite{Anninos:2008fx}.

The black hole metric (\ref{swbh}) and the warped AdS$_3$ (\ref{wads3}) are related to each other by local coordinate transformations
\bea
&& \t=-\arctan \lt[ \f{\sr{r^2-r_0^2}}{r}\sinh \lt( \f{\b\g r_0}{\sr{1-\b^2}}\phi \rt) \rt] ,\nn\\
&& \s=\arcsinh \lt[ \f{\sr{r^2-r_0^2}}{r_0}\cosh \lt( \f{\b\g r_0}{\sr{1-\b^2}}\phi \rt) \rt],  \nn\\
&& u=\b\g\sr{1-\b^2}(t-\O \phi)
     -\arctanh \lt[ \f{r}{r_0} \coth \lt( \f{\b\g r_0}{\sr{1-\b^2}}\phi \rt) \rt].
\eea
For the fast-rotating black hole one could show that
\be
-\p_\phi=\pi (T_L^J J_2 -T_R^J \td J_2),
\ee
with $T_{R,L}^J$ being these in (\ref{e2}). And for the slow-rotating black hole one just exchange $T_{R,L}^J$.

\subsection{CFT from ASG}

Indeed, the above right-moving central charge for the fast-rotating black hole, or the left-moving central charge for the slow-rotating black hole, could be derived using the BBC formalism as shown in \cite{Compere:2008cv,Compere:2009zj}. For the fast-rotating black hole in TMG, setting $\O=r_0=0$ in (\ref{swbh}), one may get the vacuum spacetime
\be
ds^2=-\f{\n^2+3}{4\n^2} dt^2+\f{\ell^2}{\n^2+3}\f{dr^2}{r^2}
     +r^2 \lt( d\phi-\f{dt}{r} \rt)^2.
\ee
There is a set of boundary conditions \cite{Compere:2007in,Compere:2009zj,Blagojevic:2009ek}
\be \label{e43}
\d g_{\m\n}=\ma O \lt(
\ba{ccc}
\f{1}{r}&\f{1}{r^2}&1\\
&\f{1}{r^3}&\f{1}{r}\\
&&r
\ea \rt)
\ee
which admit nontrivial ASG. The ASG  contains the subgroup
\be \label{e34}
\xi_m=-e^{-im\phi}(\p_\phi+mr\p_r).
\ee
with which we could got
\be
c_R=\f{(5\n^2+3)\ell}{G\n(\n^2+3)}.
 \ee
One could get the left-moving central charge using Sugawara construction as in \cite{Blagojevic:2009ek}. It is
 \be
 c_L=\f{4\n\ell}{G(\n^2+3)}.
 \ee
The central charges $c_{R,L}$ are exactly the ones obtained from the thermodynamics method. And they satisfy the diffeomorphism anomaly relation \cite{Kraus:2005zm,Solodukhin:2005ah,Anninos:2008fx}
\be
c_L-c_R=-\f{\ell}{G\n}.
\ee

For the spacelike warped black hole in NMG, we set $\O=r_0=0$ in (\ref{swbh})
\be
ds^2=-\b^2 dt^2+\f{dr^2}{\g^2 r^2}+r^2 \lt( d\phi-\f{dt}{r} \rt)^2. \label{warpvacuum}
\ee
We can check that with the same asymptotic boundary conditions (\ref{e43}) there is still ASG with the subgroup (\ref{e34}), with which we could got
\be
c_R=\f{24}{Gm\b^2\sr{2(21-4\b^2)}}.
\ee
Because of diffeomorphism invariance, there must be $c_L$ with the same value. This agrees
exactly with the ones obtained in (\ref{warpNMG}). 

\section{Null warped black hole}\label{s6}

\subsection{Black hole solution}

The null warped black hole has the metric \cite{Anninos:2008fx}
\be \label{nbh}
{ds^2}=-\frac{r^2}{r^2+\ell r+\ell^2 \a^2}dt^2
       +\frac{\ell^2}{4r^2}dr^2
       +(r^2+\ell r+\ell^2 \a^2)(d\phi-\frac{rdt}{r^2+\ell r+\ell^2 \a^2})^2,
\ee
with $\ell>0$ being the parameter of the theory and $\a>0$ being the parameter of the black hole. On construction the null warped black hole is an extremal black hole with the horizon locates at $r=0$. The black hole could be got from the slow-rotating spacelike warped black hole (\ref{swbh}) by setting
\be \label{e42}
\O=\f{\ell}{2}, ~~~ r_0^2=(1-\b^2)\a^2\ell^2,
\ee
and taking the limit $\b  \to 1$ \cite{Anninos:2008fx}. All the quantities for the null warped black hole could be got by taking proper limits of those for the spacelike warped black hole. Here it will be brief without much details.

The null warped black hole is a solution of TMG (\ref{tmg}) with $\L=-\f{1}{\ell^2}$ and $\m=-\f{3}{\ell}$. Its energy, angular momentum and entropy are respectively
\be
M=0, ~~~ J=-\f{\a^2\ell}{3G}, ~~~ S_+=\f{2\pi\a\ell}{3G}.
\ee
 Similarly the null warped black hole is a solution of NMG (\ref{nmg}) with $\l=-\f{35}{34\ell^2}$ and $m^2=\f{17}{2\ell^2}$. Its energy, angular momentum and entropy are respectively
\be
M=0, ~~~ J=-\f{4\a^2\ell}{17G}, ~~~ S_+=\f{8\pi\a\ell}{17G}.
\ee
The Hawking temperature and angular velocity of the null warped black hole is trivial $T_+=\O_+=0$, so the first law for the null warped black hole in TMG or NMG is trivially satisfied
\be
dM=T_+ d S_++\O_+ dJ.
\ee

\subsection{CFT from thermodynamics}

No matter in TMG or NMG, we always have
\bea
&& dJ=-T_L^J dS_+,  \nn\\
&& S=\f{\pi^2}{3}c_L^J T_L^J,
\eea
from which we get for TMG
\be
T_L^J=\f{\a}{\pi}, ~~~ c_L^J=\f{2\ell}{G},
\ee
and for NMG
\be
T_L^J=\f{\a}{\pi}, ~~~ c_L^J=\f{24\ell}{17G}.
\ee
Note that the CFT temperatures in TMG and NMG are the same in the null black hole case.

\subsection{CFT from ASG}

Using the boundary conditions and ASG in \cite{Anninos:2010pm}
\be
\d g_{\m\n}=\ma O \lt(
\ba{ccc}
\f{1}{r}&\f{1}{r^2}&1\\
&\f{1}{r^3}&\f{1}{r}\\
&&1
\ea \rt),
\ee
we could get a ASG still permits the subgroup (\ref{e34}), with which the central charge of the null spacetime in TMG can be got as
\be
c_L=\f{2\ell}{G},
\ee
and in NMG
\be
c_L=\f{24\ell}{17G}.
\ee
These are exactly the ones obtained by the thermodynamics method in the previous subseciton.

\section{Conclusion and discussion}\label{s7}

In this paper we used the thermodynamics method  to set up holographic pictures for the BTZ black hole, the spacelike and null warped AdS$_3$ black holes in 3D TMG and NMG. Without imposing any condition on the entropy product, we worked on the first laws of the outer and inner horizons directly, and read the universal properties of the dual CFT.  On the other side, we tried to establish a brute-force realization of the BBC formalism without calculating the superpotential by hand explicitly. We computed the masses and the angular momenta of the black holes, as well as the central charges of the global AdS$_3$ and warped AdS$_3$ spacetime in various 3D gravities in the BBC formalism. In all the cases, we found consistent agreements.

The effectiveness of the thermodynamics method for the BTZ and the warped black holes in TMG and NMG is remarkable. It could be expected that this method may be effective for other black holes in 3D higher curvature gravity theories.  In fact there are uncharged and charged black hole solutions for Born-Infeld extended new massive gravity, and their CFT duals  were also proposed \cite{Ghodsi:2010ev,Ghodsi:2011ua}. It would be nice to see if the thermodynamics method and the ASG analysis using our Mathematica code could be applied for these  black holes. It is even more interesting to see if the same is true in higher dimensional gravity.

The 3D gravity theories are special. In the cases we discussed in the present work, the black holes could be obtained from global spacetime via discrete quotient identification. As a result, the central charges of the dual CFTs can be obtained from the ASG analysis of the global spacetime. It is not necessary to take near-horizon limit. Consequently, the central charges depend on the parameters in the theories, the cosmological constant and the coupling constant, not on the hairs of the black holes. On the contrary  for the black holes in the higher dimensions, no matter rotating one or charged one, the central charges of their CFT duals are the functions of quantized charges. Moreover,  we show that even if the entropy product is mass-dependent in 3D TMG gravity, the first laws of the thermodynamics could still be applied. It seems that the first laws are more fundamental and the mass-independence condition of the entropy product could be neglected. Naively we may try to apply this philosophy to the study the AdS black holes in dimensions $d \geq 4$, which have only two physical horizons as well. It is true that the first laws allow us to read the temperatures and central charges of dual CFT. However, the central charges in the left- and right-moving sectors are different \cite{Chen:2012mh,Chen:2012ps}. This seems strange to us, and it would be nice to have better understanding for this issue.

\appendix

\section{A brute-force realization of BBC formalism}\label{sa}

The ASG analysis in \cite{Brown:1986nw,Brown:1986ed} is in the Hamiltonian formalism. While in the so-called Barnich-Brandt-Compere (BBC) formalism \cite{Barnich:2001jy,Barnich:2003xg,Barnich:2004uw,Barnich:2007bf,Compere:2007az} one has covariant systematic techniques of treating ASG based on the Lagrangian of the theory. However, for a general gravity theory other than the Einstein gravity the use of the BBC formalism is quite tedious.  One has to calculate the so-called superpotential with much labor. In the present work, we adopt a new strategy of using the BBC formalism. In fact we write a Mathematica code to implement the BBC formalism in a brute-force way. The brute-force realization of the BBC formalism is powerful especially for the higher derivative gravity when the calculation of the superpotential becomes formidable.
The superpotential for the TMG has been given in \cite{Compere:2008cv} to calculate the central charge $c_R$ of the spacelike warped black hole proposed in \cite{Anninos:2008fx}, although there are some ambiguities in some terms.\footnote{We thank St\'ephane Detournay for discussing us this subtlety.}
There was superpotential given in \cite{Nam:2010ub} of Killing vector under general background for NMG, with which one could calculate the mass and angular momentum of a black hole. However, no general superpotential formula for an asymptotic Killing vector for NMG has been given, and so the ASG analysis cannot be done.
Using our code, one could use the BBC formalism to NMG without knowing the explicit form the superpotential. Potentially, given a computer as powerful as needed, one could calculate the mass and the angular momentum of any black hole in any gravity theory in any dimensions, and read the central charges of the vacuum from ASG analysis provided appropriate asymptotic boundary conditions.

The BBC formalism is a powerful tool of calculating the charges of the asymptotic symmetry generators. For a gravity theory in $d$-dimensional spacetime, the formalism goes as follows. Given the action $I$ of the theory, one may define
\be
E_{\m\n}[g]=\f{\d I}{\d g^{\m\n}},
\ee
with $g_{\m\n}$ as the spacetime metric. For a classical background $\bar g_{\m\n}$, which is of course the solution of the equation of motion, we have $E_{\m\n}[\bar g]=0$. Expanding the metric around the background $g_{\m\n} =\bar g_{\m\n}+h_{\m\n}$, one can get the linear contribution $E^{(1)}_{\m\n}[g;\bar g]$. The asymptotic conservative current corresponding to the asymptotic Killing vector $\xi^\m$ is defined as
\be S^{\m}_\xi[h;\bar g]=E^{(1)\m\n}[g;\bar g]\xi_\n. \ee
Note that we always use the background metric $\bar g_{\m\n}$ or its inverse $\bar g^{\m\n}$ to lower or upper an index. The superpotential $k^{\m\n}_\xi[h;g]$ can be calculated as
\bea \label{e1}
&&k^{\n\m}_\xi[h;g]=\f{1}{2}\phi^i \f{\p S^{\m}_\xi}{\p\phi^i_\n}
                    +\lt( \f{2}{3}\phi^i_\a-\f{1}{3}\phi^i\p_\a \rt)\f{\p S^{\m}_\xi}{\p\phi^i_{\a\n}}
                    +\lt( \f{3}{4}\phi^i_{\a\b}-\f{1}{2}\phi^i_\a\p_\b+\f{1}{4}\phi^i\p_{\a\b} \rt)\f{\p S^{\m}_\xi}{\p\phi^i_{\a\b\n}}  \nn\\
&&\phantom{\sr{-g}k^{\m\n}=}
             +\lt( \f{4}{5}\phi^i_{\a\b\g}-\f{3}{5}\phi^i_{\a}\p_{\b\g}+\f{2}{5}\phi^i_{\a}\p_{\b\g}-\f{1}{5}\phi^i\p_{\a\b\g} \rt)
              \f{\p S^{\m}_\xi}{\p\phi^i_{\a\b\g\n}} + \cdots -(\n\lra\m),
\eea
where the field $\phi^i=h_{\r\s}$, $\phi^i_\n=h_{\r\s,\n}=\p_\n h_{\r\s}$, $\phi^i_{\a\b}=h_{\r\s,\a\b}=\p_{\a\b}h_{\r\s}=\p_\a\p_\b h_{\r\s}$, and so forth. The only nonvanishing derivatives are defined generally as
\be
\f{\p h_{\a\b,\n_1\cdots\n_k}}{\p h_{\g\d,\m_1\cdots\m_k}} \equiv \d_{\a\b}^{\g\d}\d_{\n_1\cdots\n_k}^{\m_1\cdots\m_k},
\ee
with
\be
\d_{\n_1\cdots\n_k}^{\m_1\cdots\m_k} \equiv \d_{\lt(\n_1\rt.}^{\m_1}\cdots \d_{\lt.\n_k\rt)}^{\m_k}.
\ee
The derivatives should be calculated with care, because one has to take into account the symmetries of both the indexes of the metric and the indexes of derivatives, for example,
\be
\f{\p h_{11,12}}{\p h_{11,12}}=\f{1}{2}, ~~~
\f{\p h_{12,12}}{\p h_{12,12}}=\f{1}{4}, ~~~
\f{\p h_{11,123}}{\p h_{11,123}}=\f{1}{6}, ~~~
\f{\p h_{12,123}}{\p h_{12,123}}=\f{1}{12}.
\ee
Note that for GR only the first two terms of the right hand side of (\ref{e1}) are used, for TMG the first three terms contribute, and for NMG one have to use all the four terms written out explicitly. The charge difference between $g_{\m\n}=\bar g_{\m\n}+h_{\m\n}$ and the background $\bar g_{\m\n}$ can be calculated as
\be \label{charge}
Q_{\xi}[g;\bar g]=2\int_{\p\S} d^{d-2}x k^{tr}_\xi[h;\bar g],
\ee
where $\S$ is the spatial slice of constant time $t$, $\p\S$ is the subspace of $\S$ of constant radial coordinate $r$ with $r\ra\inf$. The Poisson bracket of the charges could be calculated as
\be
\{Q_{\xi},Q_{\eta}\}=Q_{[\xi,\eta]}+K_{\xi,\eta},
\ee
where $[\xi,\eta]$ denotes the Lie derivative of the two asymptotic Killing vectors. The central charge term is
\be \label{cc}
K_{\xi,\eta} \equiv Q_{\eta}[\bar g+\ma L_\xi \bar g;\bar g]=2\int_{\p\S} d^{d-2}x k^{tr}_\eta[\ma L_\xi \bar g;\bar g].
\ee

Usually, to use the BBC formalism one needs to do three steps. Firstly, one expands the equations of motion around some background, secondly one has to calculate the supercharge $k^{\n\m}_\xi$ using (\ref{e1}) or performing integrations by parts, and lastly one calculates the charges or the central charge using (\ref{charge}) or (\ref{cc}). The popular way in using the BBC formalism is doing the first two steps by hand and then doing the last integration with Mathematica code. However this limits the application of the BBC formalism, because it is usually hard to do first two steps, especially the second one for the higher derivative gravity. We find that it is not necessary to compute the superpotential (\ref{e1}) by hand. In fact all the three steps could be done by the computer using the Mathmatica code. In this procedure, there is no ambiguity. The code can be downloaded at \url{https://s.yunio.com/Mtus0z} or \url{http://pan.baidu.com/s/1mToFv}.

\acknowledgments
The work was in part supported by NSFC Grants No. 10975005, and No. 11275010. JJZ was also in part supported by the Scholarship Award for Excellent Doctoral Student granted by the Ministry of Education of China. DLZ was partly supported by the Undergraduate Research Fund of Education Foundation of Peking University .

\providecommand{\href}[2]{#2}\begingroup\raggedright\endgroup




\end{document}